\documentclass[twocolumn,aps,nofootinbib]{revtex4}

\usepackage{graphicx}
\usepackage{epstopdf}
\usepackage{latexsym}
\usepackage{amsmath}
\usepackage{amssymb}
\usepackage{color}
\usepackage{mathrsfs}
\usepackage{lipsum}

\usepackage[center]{subfigure}

\begin{document}

 \newcommand{\bq}{\begin{equation}}
 \newcommand{\eq}{\end{equation}}
 \newcommand{\bqn}{\begin{eqnarray}}
 \newcommand{\eqn}{\end{eqnarray}}
 \newcommand{\nb}{\nonumber}
 \newcommand{\lb}{\label}
 \newcommand{\be}{\begin{equation}}
\newcommand{\en}{\end{equation}}
\newcommand{\PRL}{Phys. Rev. Lett.}
\newcommand{\PL}{Phys. Lett.}
\newcommand{\PR}{Phys. Rev.}
\newcommand{\CQG}{Class. Quantum Grav.}

\title{Quasinormal Modes by Improved Matrix Method and Weighted Residual Method}

\author{Kai Lin}\email{lk314159@hotmail.com}

\affiliation{Hubei Subsurface Multi-scale Imaging Key Laboratory, School of Geophysics and Geomatics, China University of Geosciences, Wuhan 430074, Hubei, China}

\date{\today}

\begin{abstract}
This work discusses the Improved Matrix Method and Weighted Residual Method for studying the quasinormal modes (QNMs) of black holes. In the first method, by utilizing Jordan decomposition, the improved matrix method avoids the calculation of inverse matrices in the original matrix method. This significantly saves computational resources and time needed to compute the coefficient square matrices of the derivatives.
In the second method, we illustrate the effectiveness of the weighted residual method in QNMs calculations by using the collocation method and Galerkin method as examples.
We compared the results of both methods with the Generalized Horowitz-Hubeny Method and the WKB approximation. The methods proposed in this paper demonstrates superior precision.
Additionally, we provide five criteria to exclude extraneous roots from the methods.
\end{abstract}


\maketitle

\section{Introduction}
\renewcommand{\theequation}{1.\arabic{equation}} \setcounter{equation}{0}
In 2017, the Nobel Prize Committee awarded the Nobel Prize in Physics to Rainer Weiss, Kip Thorne, and Barry Barish of the Laser Interferometer Gravitational-Wave Observatory (LIGO), a large-scale scientific project, for their outstanding contributions to the direct detection of gravitational waves. This major astronomical discovery not only proves the existence of gravitational waves but also confirms the existence of black holes in the universe. Till date, the LIGO and Virgo teams have successfully detected dozens of gravitational wave events, including those from black hole mergers as well as several binary neutron star mergers\cite{LIGO1,LIGO2,LIGO3,LIGO4,LIGO5,LIGO6,LIGO7,LIGO8,LIGO9}. The entire process of gravitational wave production by binary black holes consists of three stages. During the first stage, the binary black holes revolve around each other and generate gravitational waves which carry away their potential energy. As a result, the distance between the black holes keeps reducing, and this stage is known as the inspiral phase. The second stage is the merging of the two black holes, which generates extremely strong gravitational waves. This phase is called the merger phase. In the final stage, the newly formed black hole takes time to stabilize and go through oscillations, which is referred to as the ringdown phase. Generally, the gravitational waves during the inspiral phase are the most regular and longest-lasting of the gravitational wave signals currently detected. Calculations for the gravitational waves in this phase use the post-Newtonian approximation. The gravitational waves in the merger phase have large amplitudes but are short-lived and irregular in duration. They, therefore, require solving using numerical relativistic methods. The gravitational waves of the ringdown phase, known for their small amplitudes, are yet to be detected due to the current limitations with the gravitational wave detection accuracy. The black hole's vibration during the ringdown phase reflects many properties of the black hole itself. Therefore, ringdown gravitational waves hold rich physical connotations and offer significant value for research in theoretical physics and astrophysics. The next-generation gravitational wave detectors are likely to be precise enough to detect ringdown gravitational waves, so the study of black hole perturbation theory is one of the current areas of research interest.

After giving a gravitational perturbation in a black hole spacetime, the black hole will vibrate and radiate gravitational waves, and the process is useful for understanding gravitational wave dynamics in the ringdown phase. For an unstable black hole, analyzing the waveforms of gravitational waves generated by perturbed black holes reveals that if small perturbations cause the black hole’s gravitational vibration to grow larger and larger, and the spacetime eventually collapses. These unstable black holes cannot exist in the universe for an extended period because perturbations exist everywhere and at all times. In contrast, the gravitational vibrations of stable black holes weaken over time due to the energy carried away by gravitational waves, and eventually, the black hole’s spacetime returns to a stable state. This process typically consists of three phases: During the initial perturbation phase, the gravitational waveform of the black hole spacetime is related to different factors, making the waveform characteristics unpredictable. In the subsequent quasinormal modes (QNMs) oscillation phase, the gravitational waves exhibit distinct intrinsic vibration characteristics that conform to an eigen equation. The frequency of the gravitational wave during this stage can be described by a complex number. The real part of this complex frequency represents the frequency of the gravitational wave, and the imaginary part represents the decay rate of the gravitational wave amplitude. The vibrational characteristics of this phase depend solely on the black hole’s properties, making it useful to detect and explore the physical properties of the black hole. Finally, the gravitational wave amplitude decreases rapidly until it vanishes, and marks the phase known as late time tails: This phase’s nature also relies solely on the black hole’s properties.

Among these three phases, the oscillation phase of QNMs usually lasts the longest and is the most obvious phase that reflects the nature of the black hole. Therefore, it is also the most valuable phase to study. The physical properties in this phase can generally be described by a second-order linear differential equation. However, due to the complexity of the curved spacetime, directly calculating the analytical solution is challenging. As a result, various numerical methods have been proposed to study these problems.

In the time-domain method, the finite difference method (FDM) is commonly used to solve the QNM equation \cite{FDM1,FDM2,FDM3,FDM4,FDM5}. In the frequency-domain method, the numerical methods for solving the QNMs equation include the continuous fraction method (CFM) \cite{CFM}, asymptotic iteration method (AIM) \cite{AIM1,AIM2,AIM3}, and Horowitz-Hubeny method (HHM) \cite{HHM}. Additionally, there are other methods such as the WKB approximation\cite{WKB1,WKB2,WKB3,WKB4,WKB5,WKB6} and the P\"{o}schl-Teller potential method.

In 2016, we proposed the matrix method (MM) \cite{Lin1,Lin2,Lin3,Lin4,Lin5,Lin6,Lin7,Lin8}. This method can be widely applied to the study of QNMs in various background spacetimes, such as asymptotic flat black hole spacetimes, asymptotic (Anti-)de Sitter black hole spacetimes, neutron star spacetimes, and discontinuous potential cases. What’s more, the method offers high computational accuracy, and the precision can be increased to the required level by adding computational nodes. The computational efficiency of the method can be further improved by using the secant method and segmental interpolation \cite{Lin7}. However, this method still has one shortcoming. When calculating the coefficient square matrix of the higher order derivatives, we need to solve the inverse matrix multiple times, which consumes a significant amount of computational time when dealing with a large number of nodes.

In this paper, we analyze the coefficient square matrix of higher order derivatives and find that the matrix can be reduced to a simple form through Jordan decomposition. This finding leads to the proposal of an improved matrix method, which simplifies the process of solving the QNM equation and eliminates the need for calculating the inverse matrix. We will apply this improved matrix method to study the gravitational QNMs of the Schwarzschild black hole and compare the computational results with the generalized Horowitz-Hubeny method (GHHM) and the 6-order and 12-order WKB approximations.

Additionally, this paper will also discuss the technique of calculating QNMs equations using the Weighted Residual Method (WRM). This method is formally similar to MM but operates on completely different numerical principles. To apply the method, we expand the wave function as a series with undetermined coefficients and then integrate the master equation with a weight function over the domain definition. By imposing the condition that the integral results vanish, we transform the differential equation into an eigenmatrix equation, allowing us to solve for QNMs frequency and the corresponding wave function using the same technique as MM. In this paper, we propose an efficient procedure for applying this method to the study of QNMs problems.

In Section II, we derive the equations describing gravitational perturbations of a black hole and determine the boundary conditions for the QNM equation. The improved matrix method and the Weighted Residual Method are discussed in detail in Sections III and IV, respectively. To demonstrate the effectiveness of these two methods, we apply them to calculate the gravitational QNMs of the Schwarzschild black hole spacetime in Section V. Because numerical methods such as CFM, HHM, AIM, MM, and WRM can yield both physically meaningful solutions and extraneous roots when solving problems related to black hole QNMs, we provide several technical details in Section VI to distinguish between physical and spurious roots. Finally, Section VII includes discussions and conclusions.

\section{Master Equation for Gravitational Quasinormal Modes}
\renewcommand{\theequation}{2.\arabic{equation}} \setcounter{equation}{0}

A spherically symmetrical static black hole metric is
\bqn
\lb{Metric1}
ds^2&=&-h(r)dt^2+\frac{dr^2}{f(r)}+r^2\left(d\theta^2+\sin^2\theta d\varphi^2\right).~~~
\eqn
\textcolor{black}{
Generally, in spherically symmetric metrics, $f$ and $h$ are not necessarily equal, but $f=h$ in several famous black hole metric of general relativity. Now, we temporarily express the metric in the above form, thereby deriving gravitational perturbation equations that are more general.
}

Given a gravitational perturbation in the above static black hole spacetime background, the black hole will oscillate, and the dynamic behavior of this oscillation can be described by the gravitational perturbation equation. In the case of four-dimensional spacetime, there are ten components in the gravitational field equations, making it a challenging task to solve them simultaneously. However, the symmetry in the perturbation equation allows for the division of these ten equations into axial and polar parts. By simplifying each type of equation, we can obtain two decoupled gravitational perturbation equations \cite{GQNM1, GQNM2, GQNM3, GQNM4}. After separating the variables, the radial gravitational perturbation equations are given by
\bqn
\lb{Metric2}
\sqrt{fh}\frac{d}{dr}\left(\sqrt{fh}\frac{d\Phi(r)}{dr}\right)+(\omega^2-V(r))\Phi(r)=0,~~~
\eqn
where the potential function in axial case and polar case respectively are given by
\bqn
\lb{Metric3}
V_\text{axial}&=&\frac{1}{4}\left(f''h+h''f\right)-\frac{1}{16}\left(\frac{f}{h}h'^2+\frac{h}{f}f'^2\right)\nb\\
&&+\frac{3}{8}f'h'-\frac{1}{2r}\left(h'+fh'+f'h\right)\nb\\
&&+\frac{h+3hf^2}{4r^2f}+\frac{\lambda-2}{r^2}h,\nb\\
V_\text{polar}&=&h\frac{\lambda(6+9f^2)-\lambda^2(4+3f)+\lambda^3-9f^3}{(\lambda-3f)^2r^2}.~~~
\eqn
\textcolor{black}{In Eq.(\ref{Metric3})}, $\lambda\equiv L^2+L+1$ and $L$ is the multipole number. As we all know, the metric of Schwarzschild black hole requires
\bqn
\lb{Metric4}
\textcolor{black}{h(r)=}f(r)=1-\frac{2M}{r},
\eqn
where $r_p=2M$ is the event horizon of the black hole, and $M$ is viewed as the mass of black hole. In order to determine the boundary condition at horizon and infinity, let's introduce the tortoise coordinate
\bqn
\lb{Metric5}
r_*=\int\frac{\textcolor{black}{dr}}{f(r)}=r+r_p\ln|r-r_p|.
\eqn
Classical black holes are known to only absorb matter and not emit it. In the study of QNMs, the ingoing mode from infinity is ignored. As a result, the boundary conditions outside the black hole are typically satisfied by
\bqn
\lb{Metric6}
\Phi(r\rightarrow r_p^+)&\propto& (r-r_p)^{-i\omega r_p} e^{-i\omega r},\nb\\
\Phi(r\rightarrow \infty)&\propto& (r-r_p)^{i\omega r_p} e^{i\omega r}.
\eqn
So, to cancel the singularity at the boundary, we can set the QNM equation function as $\Phi=(r-r_p)^{-i\omega r_p}r^{2i\omega r_p}e^{i\omega r}\Psi(r)$, resulting in $\Psi(r\rightarrow r_p^+)$ and $\Psi(r\rightarrow \infty)$ becoming constants.

To apply the matrix method to the QNM of the black hole spacetime, we introduce the coordinate transformations $x=1-\frac{r_p}{r}$ and $w=\omega r_p$. This transform the domain of definition for the QNM equation to $x\in[0,1]$ \textcolor{black}{because $r\ge r_p$  in the region exterior to the black hole, so} the master equation can be expressed as follows:
\bqn
\lb{Metric7}
\alpha_2(x,\omega)\Psi''(x)+\alpha_1(x,\omega)\Psi'(x)+\alpha_0(x,\omega)\Psi(x)=0, \nb\\
\eqn
where
\bqn
\lb{Metric8}
&&\text{For axial case:}\nb\\
\alpha_0(x,\omega)&=& 3 - L - L^2 + w^2 (8 - 4 x) \nb\\
&&+ 4 i w (1 - x) - 3 x,\nb\\
\alpha_1(x,\omega)&=&  1 - 4 x + 3 x^2 - 2 i w (1 - 4 x + 2 x^2),\nb\\
\alpha_2(x,\omega)&=& (1 - x)^2 x, \nb\\
&&\text{For polar case:}\nb\\
\alpha_0(x,\omega)&=&L^4 (-2 + 3 x) + L^3 (1 + 6 x) -3 L^5\nb\\
&&- L^6 - L (1 - 3 x)^2+ 9 L^2 (1 - x) x\nb\\
&&+ 4 w^2 (1 + L + L^2 - 3 x)^2 (2 - x)\nb\\
&&+ 4 i w (1 + L + L^2 - 3 x)^2 (1 - x)\nb\\
&&- 3 (1 - x + 3 x^2 - 3 x^3)\textcolor{black}{,}\nb\\
\alpha_0(x,\omega)&=& (1 + L + L^2 - 3 x)^2 (1 - 4 x + 3 x^2)\nb\\
&&-2 i w (1 + L + L^2 - 3 x)^2 (1 - 4 x + 2 x^2)\textcolor{black}{,} \nb\\
\alpha_0(x,\omega)&=& (1 + L + L^2 - 3 x)^2 (1 - x)^2 x.
\eqn

In Eq. (\ref{Metric7}), the wave \textcolor{black}{function $\Psi$ is} regular at the boundary points $x=0$ and $x=1$. To avoid division by zero and potential numerical calculation failures, we deliberately express the coefficients of the wave function terms, as well as its first and second-order derivative terms, without denominators. In the next section, we will illustrate the fundamental principles of the matrix method and introduce an improved approach to enhance computational efficiency and resource utilization.

\section{Improved matrix method}
\renewcommand{\theequation}{3.\arabic{equation}} \setcounter{equation}{0}

There are several methods available for solving the QNM equation, but each method comes with its own limitations. The WKB approximation method offers fixed computational accuracy and often fails to achieve arbitrary desired accuracies. CFM requires intricate calculations of continuous fractional recurrence of coefficients, which can be challenging to implement in complex black hole spacetime backgrounds. AIM is difficult to apply to cases of eigenvalue problems with multiple equation couplings. Additionally, the original HHM method is solely applicable to solving the Anti-de Sitter black hole case. To overcome these drawbacks and effectively compute the QNM equation, we have introduced the matrix method. This method is highly efficient and successfully mitigates the aforementioned limitations. In this section, we will provide a detailed explanation of the matrix method.

Let's first discretize $x\in[0,1]$ into $n+1$ \emph{isometric} nodes, which can be rewritten as a column matrix $X\equiv[x_0=0,x_1,\cdot\cdot\cdot,x_n=1]^T$ with a step length of $h=1/n$. We then set the function at nodes to be expressed as a column matrix $F\equiv[f(x_0),f(x_1),\cdot\cdot\cdot,f(x_n)]^T$. Additionally, the derivatives and second-order derivatives of the functions at nodes are defined as $F'\equiv[f'(x_0),f'(x_1),\cdot\cdot\cdot,f'(x_n)]^T$ and $F''\equiv[f''(x_0),f''(x_1),\cdot\cdot\cdot,f''(x_n)]^T$, respectively. The matrix method requires us to search for coefficient square matrices $M_1$ and $M_2$ that satisfy
\bqn
\lb{MM1}
F'=M_1F,~~~~F''=M_2F.
\eqn
\textcolor{black}{Or} generally, for $K$th order derivative, we have $F^{(K)}=M_KF$. In fact, we can compute the square matrix $M_K$ using the Taylor series expansion, which is given by
\bqn
\lb{MM2}
f(x_j)=\sum^{n}_{k=0}\left(\frac{(x_j-x_i)^k}{k!}f^{(k)}(x_i)\right)+{\cal O}^{n+1}.
\eqn
Let's ignore the term ${\cal O}^{n+1}$ and expand all $f(x_j)$ at the node $x_i$ ($x_j\in X$ but $j\not=i$). This leads us to the matrix equation:
\bqn
\lb{MM3}
\begin{bmatrix}
f(x_0)-f(x_i) \\
f(x_1)-f(x_i) \\
\cdot\cdot\cdot\\
f(x_{i-1})-f(x_i) \\
f(x_{i+1})-f(x_i) \\
\cdot\cdot\cdot\\
f(x_{n})-f(x_i) \\
\end{bmatrix}=
{\cal M}
\begin{bmatrix}
f'(x_i) \\
f''(x_i) \\
\cdot\cdot\cdot\\
f^{(i)}(x_i) \\
f^{(i+1)}(x_i) \\
\cdot\cdot\cdot\\
f^{(n)}(x_i) \\
\end{bmatrix},
\eqn
where ${\cal M}$ is a square matrix, given by
\bqn
\lb{MM4}
{\cal M}\equiv
\begin{bmatrix}
\frac{(x_0-x_i)^1}{1!} & \frac{(x_0-x_i)^2}{2!} & \cdot\cdot\cdot & \frac{(x_0-x_i)^n}{n!} \\
\frac{(x_1-x_i)^1}{1!} & \frac{(x_1-x_i)^2}{2!} & \cdot\cdot\cdot & \frac{(x_1-x_i)^n}{n!} \\
\cdot\cdot\cdot & \cdot\cdot\cdot & \cdot\cdot\cdot & \cdot\cdot\cdot \\
\frac{(x_{i-1}-x_i)^1}{1!} & \frac{(x_{i-1}-x_i)^2}{2!} & \cdot\cdot\cdot & \frac{(x_{i-1}-x_i)^n}{n!} \\
\frac{(x_{i+1}-x_i)^1}{1!} & \frac{(x_{i+1}-x_i)^2}{2!} & \cdot\cdot\cdot & \frac{(x_{i+1}-x_i)^n}{n!} \\
\cdot\cdot\cdot & \cdot\cdot\cdot & \cdot\cdot\cdot & \cdot\cdot\cdot \\
\frac{(x_n-x_i)^1}{1!} & \frac{(x_n-x_i)^2}{2!} & \cdot\cdot\cdot & \frac{(x_n-x_i)^n}{n!} \\
\end{bmatrix}.
\eqn
Therefore, we can calculate the derivatives and higher-order derivatives of the function at the node $x_i$ by using the following relation:
\bqn
\lb{MM5}
\begin{bmatrix}
f'(x_i) \\
f''(x_i) \\
\cdot\cdot\cdot\\
f^{(i)}(x_i) \\
f^{(i+1)}(x_i) \\
\cdot\cdot\cdot\\
f^{(n)}(x_i) \\
\end{bmatrix}
=
\left({\cal M}^{-1}\right)
\begin{bmatrix}
f(x_0)-f(x_i) \\
f(x_1)-f(x_i) \\
\cdot\cdot\cdot\\
f(x_{i-1})-f(x_i) \\
f(x_{i+1})-f(x_i) \\
\cdot\cdot\cdot\\
f(x_{n})-f(x_i) \\
\end{bmatrix}.
\eqn
This means that we can express $F'$ and $F''$ using the elements of $F$. Clearly, the elements in the first and second rows of ${\cal M}^{-1}$ correspond to the elements in the $i-$th row of $M_1$ and $M_2$ respectively, as shown in Eq.(\ref{MM1}).

For an eigenequation with an eigenvalue $\omega$ and an eigenfunction $f(x)$
\bqn
\lb{MM6}
\alpha_2(x,\omega)f''(x)+\alpha_1(x,\omega)f'(x)+\alpha_0(x,\omega)f(x)=0,~~~~~
\eqn
By substituting $f'(x)$ and $f''(x)$ with the column matrices $F'=M_1F$ and $F''=M_2F$ respectively, the eigenequation can be rewritten as a matrix equation
\bqn
\lb{MM7}
A_2M_2F+A_1M_1F+A_0F=0,
\eqn
where $A_i\equiv\text{diag}\left[\alpha_i(x_0,\omega),\alpha_i(x_1,\omega),\ldots,\alpha_i(x_n,\omega)\right]$ is a square diagonal matrix. Thus, the eigenvalue $\omega$ must satisfy the condition that the determinant of the matrix vanishes:
\bqn
\lb{MM8}
\det(A_2M_2+A_1M_1+A_0)=0.
\eqn

Due to the fact that the neglected term is of ${\cal O}^{n+1}$ order, the matrix method achieves a high level of precision. Furthermore, in the works\cite{Lin1,Lin2,Lin3,Lin4,Lin5,Lin6,Lin7}, we have demonstrated the applicability of this method in calculating QNM frequencies across various spacetime scenarios. However, the derivation mentioned above reveals that the computation of the inverse matrix ${\cal M}^{-1}$ needs to be performed multiple times, resulting in significant consumption of computational resources and time. Taking these factors into account, the \emph{improved matrix method} is proposed.

Now, let's analyze the column matrix $M_i$. Since $F''=M_1F'=M_1M_1F=M_1^2F$, according to Eq.(\ref{MM1}), we have $M_2=M_1^2$. Generally, the column matrix of the $k$-th order derivative $F^{(k)}$ should satisfy
\bqn
\lb{MM9}
F^{(k)}=M_kF=M_1^kF,
\eqn
so we have $M_k=M_1^k$. This means that we don't need to calculate $M_k$ individually; instead, we only need to consider the power of $M_1$.

Next, We will conduct a Jordan decomposition of $M_1$:
\bqn
\lb{MM10}
M_1=P_KJ_KP_K^{-1},
\eqn
where $P_K$ is a reversible square matrix, which elements is given by
\bqn
\lb{MM11}
P_K\equiv\left[\left(P_K\right)_{ij}\right]\equiv\left[\frac{1}{(j-1)!}\left(\frac{i-n-1}{n}\right)^{j-1}\right],~~~~~
\eqn
It should be noted that the element at row $i=n+1$ and column $j=1$ is an indeterminate $0^0$, but we set this element to be $1$ in matrix $P_K$. 

\textcolor{black}{
On the other hand, due to the special property of $M_1$, all values on the diagonal of the matrix $J_K$ vanish. Hence, the nilpotent Jordan matrix $J_K$ is precisely a nilpotent Jordan block, which is given by:
}
\bqn
\lb{MM12}
J_K\equiv\left[\left(J_K\right)_{ij}\right]\equiv\left[\delta_{i,j-1}=
\left\{\begin{matrix}
1 & \text{as~}i=j-1\\
0 & \text{as~}i\not=j-1\\
\end{matrix}\right.\right].~~~~~
\eqn

What’s more, \textcolor{black}{$J_K$ possesses} an interesting property, which can be stated as follows:

\bqn
\lb{MM13}
J_K^k=\left[\delta_{i,j-k}=
\left\{\begin{matrix}
1 & \text{as~}i=j-k\\
0 & \text{as~}i\not=j-k\\
\end{matrix}\right.\right], k=0,1,2,\cdot\cdot\cdot.~~~~~~~
\eqn
Because all eigenvalues of square matrices $M_k$ (where $k\ge1$) are zero, these matrices are not invertible, and the diagonal elements of $J_K^k$ (where $k\ge1$) are all zero. We define $J_K^0$ as the identity matrix $I$. According to Equation (\ref{MM10}), we obtain
\bqn
\lb{MM14}
M_k=P_KJ_K^kP_K^{-1}, k=0,1,2,\cdot\cdot\cdot.
\eqn
Substituting Eq.(\ref{MM9}) into Eq.(\ref{MM7}), we have
\bqn
\lb{MM15}
\mathbb{M}\mathbb{F}&=&0,\nb\\
\mathbb{M}&\equiv& A_2P_KJ_KJ_K+A_1P_KJ_K+A_0P_K,\nb\\
\mathbb{F}&\equiv& \left(P_K^{-1}\right)F,
\eqn
and the QNM frequency, as an eigenvalue, can be obtained by solving the determinant equation
\bqn
\lb{MM16}
\det\left(A_2P_KJ_K^2+A_1P_KJ_K+A_0P_K\right)=0.
\eqn
After obtaining the eigenvalues, we can directly calculate the eigenvector matrix $\mathbb{F}$ (for example, this calculation can be accomplished using the \emph{NSolve} command or the \emph{FindRoot} command in \emph{MATHEMATICA}), and finally obtain the results of the function matrix $F$ using the following relation
\bqn
\lb{MM17}
F=P_K\mathbb{F}.
\eqn

From Eq. (\ref{MM16}) and Eq. (\ref{MM17}), it is evident that the computational effort in solving the QNM problem is significantly reduced, as we avoid calculating the inverse matrix in this approach. In Sec. IV, we will discuss another effective numerical method, known as the Weighted Residual Method.

\section{Weighted Residual Method}
\renewcommand{\theequation}{4.\arabic{equation}} \setcounter{equation}{0}

The matrix method is essentially a high-precision difference method that transforms a differential equation into a matrix equation by expressing the derivatives at the $i$-th node in terms of numerical values at all nodes. Then, we can calculate the QNM frequency and wave function as eigenvalues and eigenvectors of the matrix equation. Therefore, in principle, the determination of square matrices $M_K$ can also be achieved using other numerical differentiation methods such as the Differential Quadrature Method or the Runge-Kutta method. In fact, there are various other numerical methods for solving differential equations in computational physics. One well-known method is the Weighted Residual Method (WRM), which plays a significant role, especially in finite element methods. In this section, we discuss the application of this method for solving QNM problems.

To apply the Weighted Residual Method (WRM), a transformation is performed as follows:
\bqn
\lb{WRM1}
\Psi(x)=\frac{\Xi(x)}{x(1-x)},
\eqn
so that $\Xi$ satisfies the condition
\bqn
\lb{WRM2}
\Xi(x=0)=\Xi(x=1)=0,
\eqn
and Eq.(\ref{Metric7}) becomes
\bqn
\lb{WRM3}
\beta_2(x,\omega)\Xi''(x)+\beta_1(x,\omega)\Xi'(x)+\beta_0(x,\omega)\Xi(x)=0,\nb\\
\eqn
where
\bqn
\lb{WRM4}
\beta_0(x,\omega)&=&2(3x^2-3x+1)\alpha_2(x,\omega)\nb\\
&&+x(1-x)(2x-1)\alpha_1(x,\omega)\nb\\
&&+x^2(1-x)^2\alpha_0(x,\omega),\nb\\
\beta_1(x,\omega)&=&2x(1-x)(2x-1)\alpha_2(x,\omega)\nb\\
&&+x^2(1-x)^2\alpha_1(x,\omega),\nb\\
\beta_2(x,\omega)&=&x^2(1-x)^2\alpha_2(x,\omega).
\eqn

According to the idea of WRM, we need to provide a trial function.
\bqn
\lb{WRM5}
\tilde{\Xi}(x)=\sum^N_{i=1}a_iu_i(x),
\eqn
where \textcolor{black}{$a_i$} is the set of undetermined coefficients, and $u_i(x)$ is a complete set of functions that satisfy the boundary condition (\ref{WRM2}). In this paper, for the sake of simplicity, we choose the trial function as follows:
\bqn
\lb{WRM6}
u_i(x)=x^i(1-x).
\eqn
\textcolor{black}{According to the Weighted Residual Method, we need to introduce a weight function $\sigma_k$.} By substituting the above trial function into Eq.(\ref{WRM3}) and integrating with \textcolor{black}{the} weight function $\sigma_k$, we obtain:

\bqn
\lb{WRM7}
I_k&\equiv&\int_0^1\sigma_k\left[\beta_2\tilde{\Xi}''+\beta_1\tilde{\Xi}'+\beta_0\tilde{\Xi}\right]dx\nb\\
&=&\sum^N_{i=1}\left[\int^1_0\sigma_k(\beta_2u_i''+\beta_1u_i'+\beta_0u_i)dx\right]a_i\nb\\
&\equiv&\sum^N_{i=1}\mathbb{I}_{ki}a_i=0
\eqn
where $k=1,2,\cdot\cdot\cdot,N$. After choosing the weight function $\sigma_k$, Eq.(\ref{WRM7}) can be reexpressed as a matrix equation with a square matrix $M_\text{WRM}$ of size $N\times N$, as follows:
\bqn
\lb{WRM8}
M_\text{WRM}F_\text{WRM}\equiv\begin{bmatrix}
\mathbb{I}_{11} & \mathbb{I}_{12} & \cdot\cdot\cdot & \mathbb{I}_{1N} \\
\mathbb{I}_{21} & \mathbb{I}_{22} & \cdot\cdot\cdot & \mathbb{I}_{2N} \\
\cdot\cdot\cdot & \cdot\cdot\cdot & \cdot\cdot\cdot & \cdot\cdot\cdot \\
\mathbb{I}_{N1} & \mathbb{I}_{N2} & \cdot\cdot\cdot & \mathbb{I}_{NN} \\
\end{bmatrix}\begin{bmatrix}
a_1 \\
a_2 \\
\cdot\cdot\cdot\\
a_N \\
\end{bmatrix}=0.~~~
\eqn
Similar to MM, the eigenvalue $\omega$ can be obtained by solving the determinant equation:
\bqn
\lb{WRM9}
\det\left(M_\text{WRM}\right)=0,
\eqn
and then, by solving for the corresponding eigenvector $F_\text{WRM}=[a_1,a_2,\cdots,N]^T$, $\tilde{\Xi}$ can be determined.

The next problem is how to choose the weight function $\sigma_k$? In the weighted residual method, common methods for selecting weight functions include the collocation method, Galerkin method, least squares method, subdomain method, method of moments, and others. In this paper, we choose the collocation method and Galerkin method as examples, to solve the QNMs' master equation.

In the collocation method, abbreviated as WRM-C, the weight function is determined as follows:
\bqn
\lb{WRM10}
\sigma_k=\delta(x-x_k),
\eqn
so that
\bqn
\lb{WRM11}
\mathbb{I}_{ki}&=&\int^1_0\delta(x-x_k)(\beta_2u_i''+\beta_1u_i'+\beta_0u_i)dx\nb\\
&=&\beta_2(x_k,\omega)u_i''(x_k)+\beta_1(x_k,\omega)u_i'(x_k)\nb\\
&&+\beta_0(x_k,\omega)u_i(x_k),
\eqn
while in the Galerkin method, abbreviated as WRM-G, the weight function is given by
\bqn
\lb{WRM12}
\sigma_k=\frac{\partial\tilde{\Xi}}{\partial a_k}=u_k(x)
\eqn
so that
\bqn
\lb{WRM13}
\mathbb{I}_{ki}=\int^1_0u_k(\beta_2u_i''+\beta_1u_i'+\beta_0u_i)dx.
\eqn
By substituting the above formulas into Eq.(\ref{WRM8}) and Eq.(\ref{WRM9}), we can determine the QNMs frequency, denoted as $w$, and the corresponding wave function, represented by $\tilde{\Xi}$.

In the next section, we will calculate the Quasinormal Modes of the Schwarzschild black hole spacetime using both the matrix method and the weighted residual method.

\section{Numerical Result}
\renewcommand{\theequation}{5.\arabic{equation}} \setcounter{equation}{0}
Using the improved matrix method presented in Section III and the weighted residual method shown in Section IV, we solve the master equation for QNMs of the Schwarzschild black hole spacetime discussed in Section II. The calculations can be efficiently performed using \emph{MATHEMATICA} software.

To demonstrate the precision of the matrix method, we simultaneously calculated the QNMs frequencies of $L=2$ overtone numbers $n=0$ of the Schwarzschild spacetime using the matrix method with 17 nodes, WRM-C with $N=14$, WRM-G with $N=14$, the Generalized Horowitz-Hubeny Method (GHHM), and the 6th- and 12th-order WKB approximations. We presented the results in TABLE I. The GHHM can be found in \cite{Lin8}, and we set the order of the method as $N=100$ in this work, which allows us to consider the GHHM results as the exact values. The 6th-order WKB approximation was first obtained by Konoplya in 2003 \cite{WKB3}, and recently, the 12th-order WKB approximation has been derived in \cite{WKB5, WKB6}. However, the higher-order WKB approximation formulas are lengthy and complicated, making it nearly impossible to manually input them into codes for orders higher than 6th. Nevertheless, the table shows that the results of the highest-order WKB approximation still have large errors, while the precision of the 17-point matrix method and weighted residual method with $N=14$ is already very close to the exact values of the QNMs frequencies. From the high-precision calculation results, we find that the frequencies of QNMs in both axial and polar cases are exactly the same \textcolor{black}{\cite{EOQNM1,EOQNM2,EOQNM3}}. However, expressing this property using the WKB approximation becomes difficult due to the existence of calculation errors.

\begin{table*}[htbp]
\caption{\label{Table1} Gravitational quasinormal mode frequency $w$ obtained by the Matrix Method (MM), Weighted Residual Method with Collocation (WRM-C), Weighted Residual Method with Galerkin (WRM-G), Generalized Harowitz-Hubeny Method (GHHM), and the WKB approximation with $n=0$.
}
\centering
\begin{tabular}{c c c c}
         \hline\hline
 &$17$ nodes MM& GHHM ($N=100$) &  $6$ order WKB    \\
        \hline
$L=2, w_\text{axial}$~~~~ &$0.747343 - 0.177925i$~~~~&$0.747343 - 0.177925i$~~~~&$0.747722 - 0.177488i$\\
$L=2, w_\text{polar}$~~~~ &$0.747343 - 0.177925i~~~~$&$0.747343 - 0.177925i$~~~~&$0.672078 - 0.101428i$\\
$L=3, w_\text{axial}$~~~~ &$1.19889 - 0.185406i$~~~~&$1.19889 - 0.185406i$~~~~&$1.19889 - 0.185406i$\\
$L=3, w_\text{polar}$~~~~ &$1.19889 - 0.185406i$~~~~&$1.19889 - 0.185406i$~~~~&$1.15671 - 0.115745i$\\
$L=4, w_\text{axial}$~~~~ &$1.61836 - 0.188328i$~~~~&$1.61836 - 0.188328i$~~~~&$1.61836 - 0.188328i$\\
$L=4, w_\text{polar}$~~~~ &$1.61836 - 0.188328i$~~~~&$1.61836 - 0.188328i$~~~~&$1.58577 - 0.124658i$\\
\hline
\hline
 &WRM-C ($N=14$)& WRM-G ($N=14$) &   $12$ order WKB    \\
        \hline
$L=2, w_\text{axial}$~~~~ &$0.747338 - 0.177924i$~~~~&$0.747324 - 0.177924i$~~~~&$0.7487 - 0.188126i$\\
$L=2, w_\text{polar}$~~~~ &$0.747339 - 0.177922i~~~~$&$0.747343 - 0.177925i$~~~~&$0.681726 - 0.122566i$\\
$L=3, w_\text{axial}$~~~~ &$1.19889 - 0.185406i$~~~~&$1.19889 - 0.185406i$~~~~&$1.19922 - 0.189856i$\\
$L=3, w_\text{polar}$~~~~ &$1.19889 - 0.185406i$~~~~&$1.19889 - 0.185406i$~~~~&$1.15664 - 0.123991i$\\
$L=4, w_\text{axial}$~~~~ &$1.61836 - 0.188328i$~~~~&$1.61836 - 0.188328i$~~~~&$1.61849 - 0.19089i$\\
$L=4, w_\text{polar}$~~~~ &$1.61836 - 0.188328i$~~~~&$1.61836 - 0.188328i$~~~~&$1.58542 - 0.129529i$\\
\hline
\hline
\end{tabular}
\end{table*}

Next, we calculated the QNMs' frequencies $w=w_\text{Re}+iw_\text{Im}$ in the black hole spacetime and presented the frequencies with smaller overtone numbers $n$ in Figure 1. The figures suggest a quasi-linear dependence of QNMs' frequency on $n$ for a fixed $L$.

\begin{figure}[htbp]
\centering
\includegraphics[width=0.8\columnwidth]{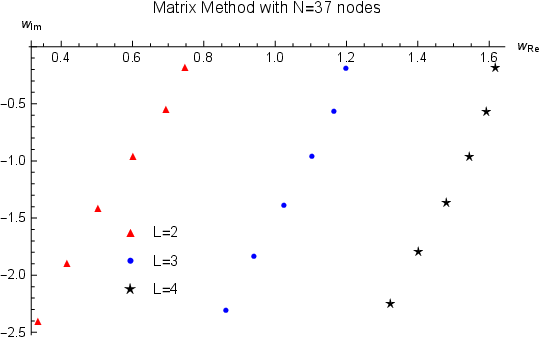}
\caption{Gravitational quasinormal mode frequencies calculated using the matrix method with 37 nodes}
\lb{Fig1}
\end{figure}
Figures 2-4 illustrate the eigenvectors (wave functions) corresponding to the eigenvalues (QNMs frequencies) calculated by the matrix method for the cases of $n=0,1$ and $L=2,3,4$ respectively. Here, due to the complicated form of the wave function $\Phi(r)$, we perform the transformation $\Phi(r)=(r-r_p)^{-i\omega r_p}r^{2i\omega r_p}e^{i\omega r}\Psi(r)$ to simplify the wave function and draw the relation between $\Psi$ and $x=1-\frac{r_p}{r}$. It can be observed that the QNMs frequencies $w$ are the same in both the axial and polar cases, but the corresponding wave functions are still different.

\begin{figure*}[htbp]
\centering
\includegraphics[width=0.8\columnwidth]{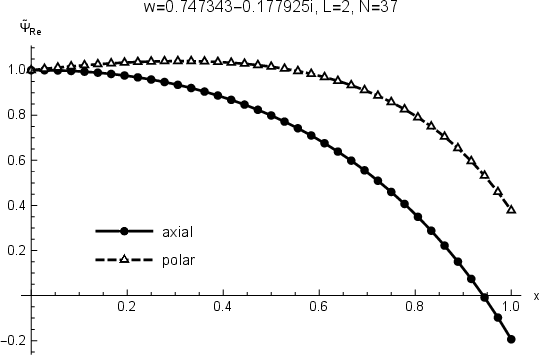}\includegraphics[width=0.8\columnwidth]{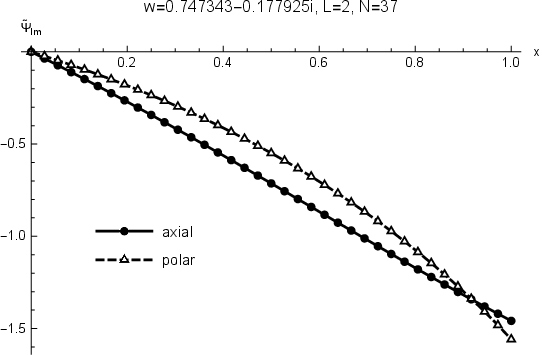}
\includegraphics[width=0.8\columnwidth]{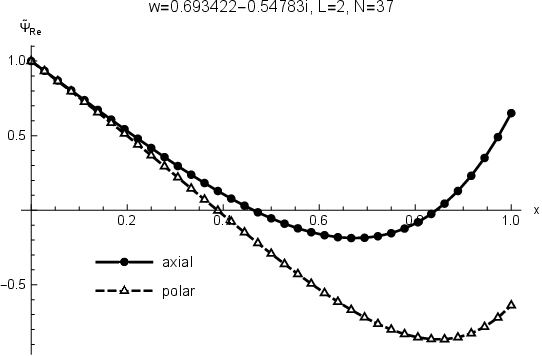}\includegraphics[width=0.8\columnwidth]{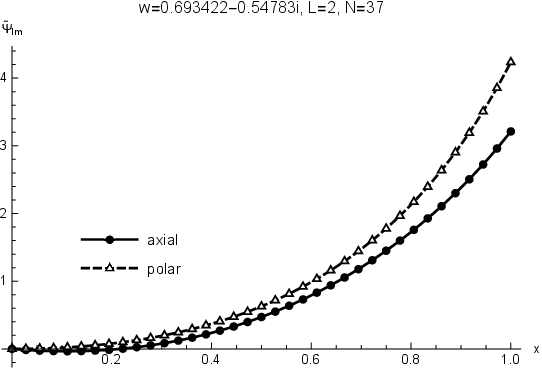}
\caption{Quasinormal mode frequencies and wave functions in the Schwarzschild black hole spacetime were obtained using the matrix method with 37 nodes. Here, we show the case of $L=2$. The wave function is transformed as $\Phi(r)=(r-r_p)^{-i\omega r_p}r^{2i\omega r_p}e^{i\omega r}\Psi(r)$, and $\tilde{\Psi}(r)\equiv\Psi(r)/\Psi(r_p)$. The variable $x$ is defined as $x=1-\frac{r_p}{r}$, and the quasinormal mode frequency is denoted as $w=\omega r_p$.
}
\lb{Fig2}
\end{figure*}

\begin{figure*}[htbp]
\centering
\includegraphics[width=0.8\columnwidth]{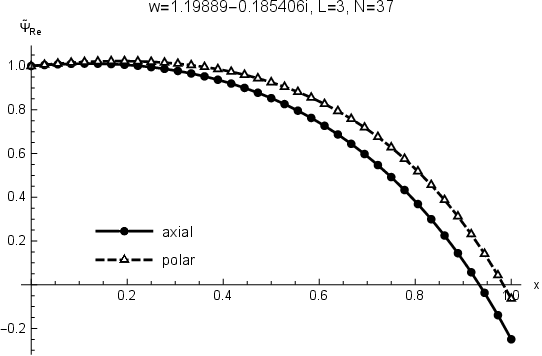}\includegraphics[width=0.8\columnwidth]{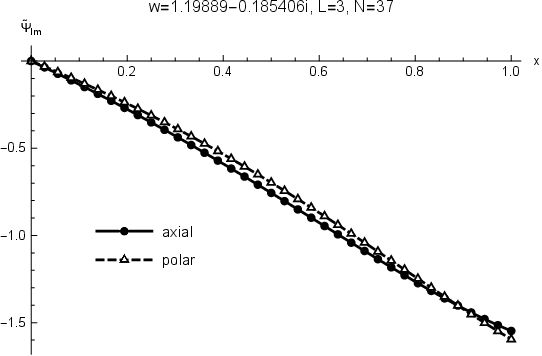}
\includegraphics[width=0.8\columnwidth]{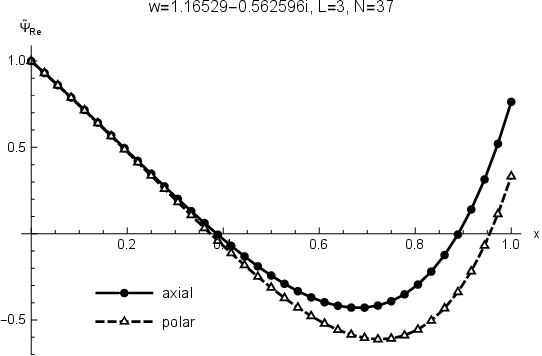}\includegraphics[width=0.8\columnwidth]{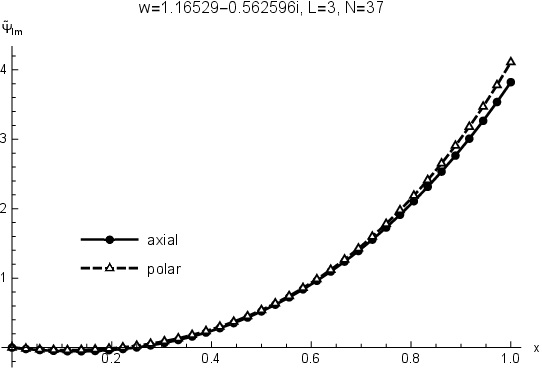}
\caption{Quasinormal mode frequencies and wave functions in the Schwarzschild black hole spacetime were obtained using the matrix method with 37 nodes. Here, we show the case of $L=3$. The wave function is transformed as $\Phi(r)=(r-r_p)^{-i\omega r_p}r^{2i\omega r_p}e^{i\omega r}\Psi(r)$, and $\tilde{\Psi}(r)\equiv\Psi(r)/\Psi(r_p)$. The variable $x$ is defined as $x=1-\frac{r_p}{r}$, and the quasinormal mode frequency is denoted as $w=\omega r_p$.
}
\lb{Fig3}
\end{figure*}

\begin{figure*}[htbp]
\centering
\includegraphics[width=0.8\columnwidth]{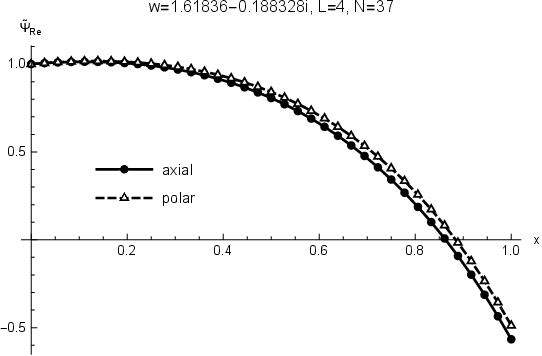}\includegraphics[width=0.8\columnwidth]{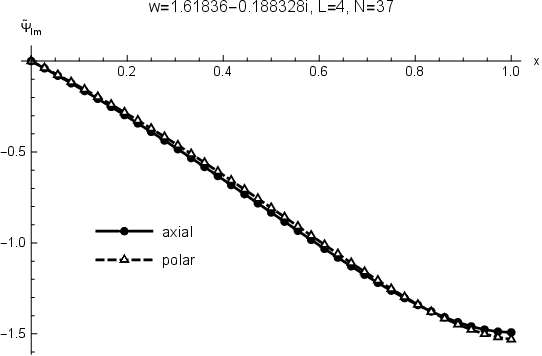}
\includegraphics[width=0.8\columnwidth]{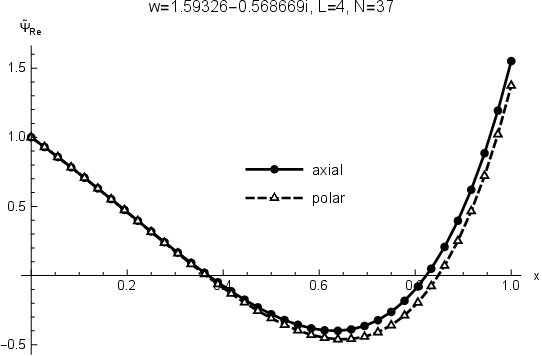}\includegraphics[width=0.8\columnwidth]{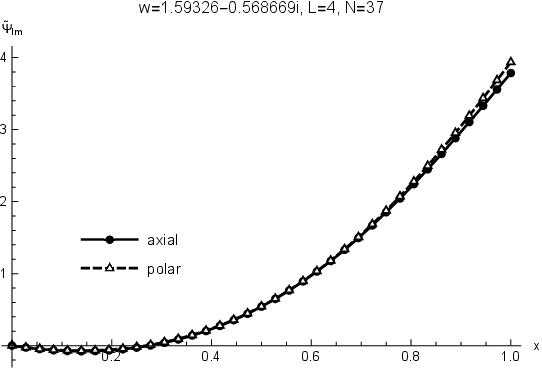}
\caption{Quasinormal mode frequencies and wave functions in the Schwarzschild black hole spacetime were obtained using the matrix method with 37 nodes. Here, we show the case of $L=4$. The wave function is transformed as $\Phi(r)=(r-r_p)^{-i\omega r_p}r^{2i\omega r_p}e^{i\omega r}\Psi(r)$, and $\tilde{\Psi}(r)\equiv\Psi(r)/\Psi(r_p)$. The variable $x$ is defined as $x=1-\frac{r_p}{r}$, and the quasinormal mode frequency is denoted as $w=\omega r_p$.
}
\lb{Fig4}
\end{figure*}

In Figure 5, we compare the results of wave functions using the matrix method and the weighted residual method for the cases of $L=2,3,4$ and $n=0$, with the wave function given by $\Phi(r)=(1-\frac{r_p}{r})^{-1}(\frac{r_p}{r})^{-1}(r-r_p)^{-i\omega r_p}r^{2i\omega r_p}e^{i\omega r}\Xi(r)$. It was found that the results of the MM, WRM-C, and WRM-G match very well.

\begin{figure*}[htbp]
\centering
\includegraphics[width=0.7\columnwidth]{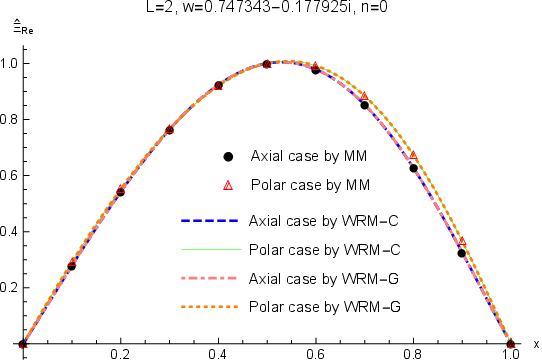}\includegraphics[width=0.7\columnwidth]{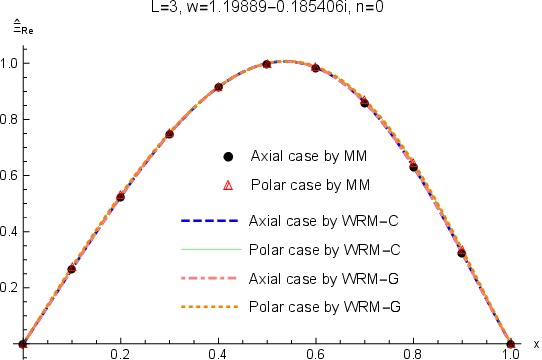}\includegraphics[width=0.7\columnwidth]{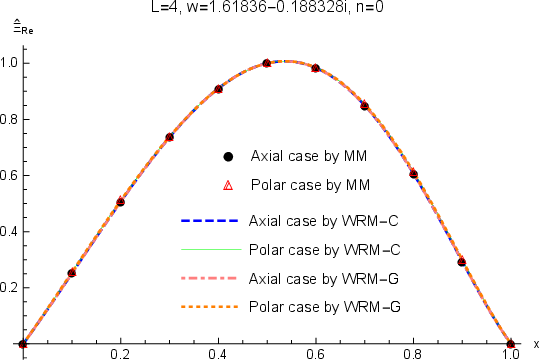}
\includegraphics[width=0.7\columnwidth]{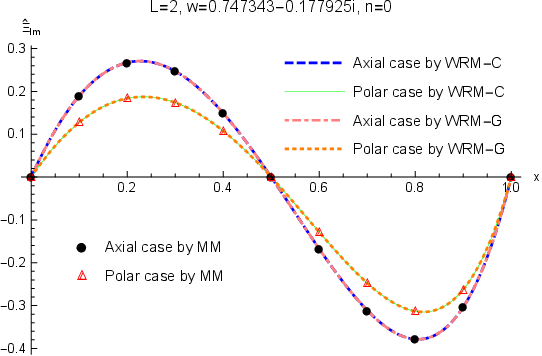}\includegraphics[width=0.7\columnwidth]{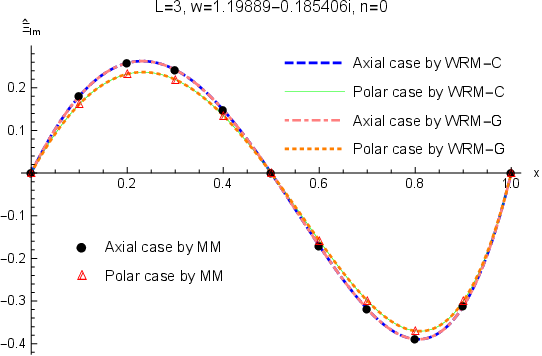}\includegraphics[width=0.7\columnwidth]{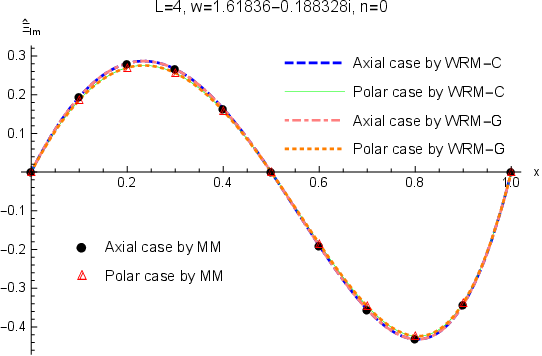}
\caption{Quasinormal mode frequencies and wave functions in the Schwarzschild black hole spacetime were obtained using the Weighted Residual Method with $N=14$. Here, we show the cases of $L=2,3,4$. The wave function is transformed as $\Phi(r)=(1-\frac{r_p}{r})^{-1}(\frac{r_p}{r})^{-1}(r-r_p)^{-i\omega r_p}r^{2i\omega r_p}e^{i\omega r}\Xi(r)$, and $\hat{\Xi}(x)\equiv\frac{\Xi(x)}{\Xi(\frac{1}{2})}$. The variable $x$ is defined as $x=1-\frac{r_p}{r}$, and the quasinormal mode frequency is denoted as $w=\omega r_p$.
}
\lb{Fig5}
\end{figure*}

In calculations using the matrix method and the weighted residual method, many roots may be obtained simultaneously for Eq.(\ref{MM8}) or Eq.(\ref{WRM9}). Some of these roots are physical solutions, while others are extraneous roots. The question of how to exclude the extraneous roots and identify the physical solutions is an unavoidable problem that also affects the application of various methods, such as CFM, HHM, and AIM. Therefore, in the next section, we will discuss how to exclude the extraneous roots.

\section{How can we exclude the extraneous roots?}
\renewcommand{\theequation}{6.\arabic{equation}} \setcounter{equation}{0}

Many numerical methods have been proposed to solve the problem of QNMs, and these methods can be grouped into three categories: 

I. Analytic-like methods, such as the Monodromy Method, P\"{o}schl-Teller Potential Method, and WKB Approximation.

II. Time-domain methods, namely FDM (Finite Difference Method).

III. Frequency-domain methods, such as CFM (Continued Fraction Method), HHM (Horowitz-Hubeny Method), AIM (Asymptotic Iteration Method), MM (Matrix Method), and WRM (Weighted Residual Method).

Analytic-like methods solve the equation by using several analytic approximate solutions, which can obtain QNMs frequencies with overtone number $n$, but the precision of the results is not high due to the approximation. For the time-domain method, or FDM, it needs to solve a partial differential equation with time coordinate $t$ and radial coordinate $r$. The results can illustrate the dynamic process from the initial perturbation phase to the QNMs oscillation phase and the late-time tails phase. From the results of FDM, we can extract the QNMs frequencies using the Prony method, but the precision is very low. Therefore, to obtain high-precision results of QNMs frequencies, we need to solve it using the frequency-domain method, which can improve the computational precision by increasing the order of the method. In theory, the results can reach any precision if the conditions of hardware and software allow.

In fact, the frequency-domain method is used to solve the eigenvalue equation obtained through Fourier transformation in the frequency domain, where the eigenvalue and the corresponding eigenfunction represent the QNMs frequency and wave function, respectively. By employing numerical techniques, we can simplify the problem of solving QNMs frequencies into solving the following nonlinear equation:
\bq
\lb{H1}
f(w)=0.
\eq
The above nonlinear equation can often provide multiple solutions, which include both meaningful physical solutions and spurious extraneous roots. Therefore, in most cases, we need to distinguish between these solutions and exclude the extraneous roots. In this section, we will use the matrix method as an example to illustrate several criteria for selecting the appropriate physical solutions for QNMs frequencies.

Criterion I: The simplest way to select the physical solutions from Eq.(\ref{H1}) is to iteratively calculate the equation with an initial value, which can be obtained through analytic-like methods or the Prony method involving FDM. For instance, we can utilize the \emph{FindRoot} command within the \emph{MATHEMATICA} software to directly apply the Newton Iteration Method for solving Eq.(\ref{H1}).

Criterion II: As the order of the frequency-domain method increases, the function $f=f(w)$ becomes more complicated, resulting in Eq.(\ref{H1}) possessing more roots. For physical roots, the QNMs frequency value converges rapidly as the order of the frequency-domain method increases. However, extraneous roots diverge as the order increases. In the \emph{MATHEMATICA} software, it is easy to find all complex roots of Eq.(\ref{H1}) using the \emph{NSolve} command. Alternatively, we can track a root and select physical roots based on the criterion. In order to demonstrate the distinction between physical solutions and extraneous roots, TABLE II lists the five roots with the smallest absolute values and $w_\text{Im}\le0$ for the 10-node to 21-node cases. From this table, we can observe the validity of the criterion.

\begin{table*}[htbp]
\caption{\label{Table} 
The five roots with the smallest absolute values and $w_\text{Im}\le0$, $w_\text{Re}\ge0$ are obtained by applying the Matrix Method from the 10-node case to the 21-node case, where $L=2$. It has been observed that both the cases of $w_{n=0}=0.747-0.178i$ and $w_{n=1}=0.693-0.548i$ exhibit convergence, indicating that they are physical solutions. Conversely, any other roots with an absolute value smaller than that of $w_{n=1}$ are considered extraneous roots. For roots with absolute values greater than $w_{n=1}$, their convergence needs to be evaluated by comparing them with the roots obtained from higher order calculations in order to determine whether they are physical solutions.
}
\centering
\begin{tabular}{c r r r r r}
         \hline\hline
10 nodes ~~~~ &$-0.468294i$~~~~&$0.747508 - 0.177933i$~~~~&$0.689288 - 0.547424i$~~~~&$0.647883 - 0.936687i$~~~~&$0.401398 - 1.30967i$\\
11 nodes ~~~~ &$-0.382939i$~~~~&$0.74727 - 0.177959i$~~~~&$0.69513 - 0.545721i$~~~~&$0.56141 - 0.987281i$~~~~&$0.553994 - 1.25564i$\\
12 nodes ~~~~ &$-0.323932i$~~~~&$0.747365 - 0.177886i$~~~~&$0.693071 - 0.549614i$~~~~&$-0.938299i$~~~~&$0.596633 - 0.939239i$\\
13 nodes ~~~~ &$-0.281291i$~~~~&$-0.743462i$~~~~&$0.747343 - 0.17795i$~~~~&$0.692912 - 0.547003i$~~~~&$0.619642 - 0.963036i$\\
14 nodes ~~~~ &$-0.24828i$~~~~&$-0.622664i$~~~~&$0.747337 - 0.177912i$~~~~&$0.694103 - 0.548085i$~~~~&$0.590402 - 0.956948i$\\
15 nodes ~~~~ &$-0.222168i$~~~~&$-0.537063i$~~~~&$0.747351 - 0.177929i$~~~~&$0.692989 - 0.547906i$~~~~&$0.605296 - 0.949344i$\\
16 nodes ~~~~ &$-0.200813i$~~~~&$-0.474578i$~~~~&$0.747338 - 0.177924i$~~~~&$0.693612 - 0.547599i$~~~~&$-1.01441i$\\
17 nodes ~~~~ &$-0.183099i$~~~~&$-0.424556i$~~~~&$0.747346 - 0.177923i$~~~~&$-0.856853i$~~~~&$0.693385 - 0.548037i$\\
18 nodes ~~~~ &$-0.168111i$~~~~&$-0.384707i$~~~~&$-0.752356i$~~~~&$0.747342 - 0.177926i$~~~~&$0.693365 - 0.547707i$\\
19 nodes ~~~~ &$-0.155299i$~~~~&$-0.351306i$~~~~&$-0.671985i$~~~~&$0.747344 - 0.177923i$~~~~&$0.693506 - 0.547884i$\\
20 nodes ~~~~ &$-0.1442i$~~~~&$-0.32338i$~~~~&$-0.607288i$~~~~&$0.747344 - 0.177925i$~~~~&$0.693355 - 0.547825i$\\
21 nodes ~~~~ &$-0.134509i$~~~~&$-0.299317i$~~~~&$-0.555206i$~~~~&$0.747343 - 0.177924i$~~~~&$0.693464 - 0.547806i$\\
\hline
\hline
\end{tabular}
\end{table*}

Criterion III: In physical solutions, eigenvalues with lower frequencies, corresponding to smaller overtone numbers $n$, converge to the QNMs frequencies earlier than eigenvalues with higher frequencies and greater overtone numbers $n$. Therefore, identifying physical solutions with smaller $n$ by increasing the order of frequency-domain methods is reliable. However, this property does not apply to extraneous roots. Figure 6 demonstrates all roots obtained using the Matrix Method under the conditions $-2\le w_\text{Im}\le0$, $1\ge w_\text{Re}\ge0$, and $L=2$. It can be observed that eigenvalues with $n=0$ and $n=1$ converge to the values of the QNMs frequencies in the 11-node case. Conversely, eigenvalues with $n=2$ and $n=3$ approach the QNMs frequencies in the 17-node case and 25-node case, respectively.

\begin{figure}[htbp]
\centering
\includegraphics[width=0.8\columnwidth]{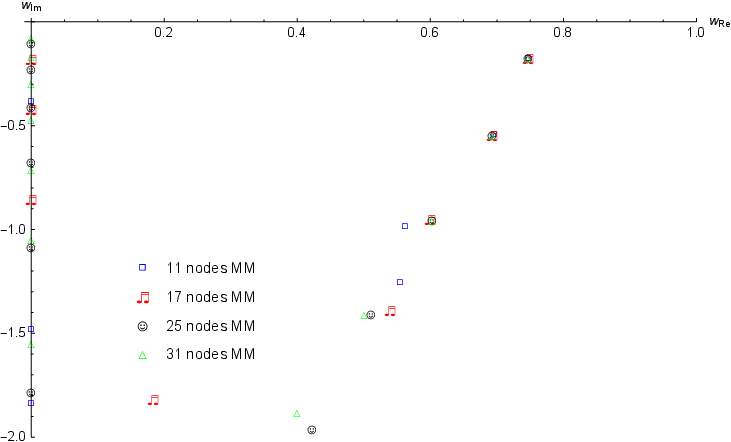}
\caption{The roots of the master equation for axial gravitational QNMs were obtained using the matrix method for cases with 11, 17, 25, and 31 nodes, respectively, where $L=2$.
}
\lb{Fig6}
\end{figure}

Criterion IV: The physical solutions obtained from different frequency-domain methods are concentrated near the QNMs frequencies, while the distribution of extraneous roots from different methods is widely scattered. Figure 7 shows the roots of the master equation for axial gravitational QNMs with $L=2$ using the matrix method, Weighted Residual Method with the collocation method, and Weighted Residual Method with the Galerkin method. It is evident that the physical roots of lower frequencies tend to cluster together across different methods. Therefore, physical solutions can be identified based on this criterion.

\begin{figure}[htbp]
\centering
\includegraphics[width=0.8\columnwidth]{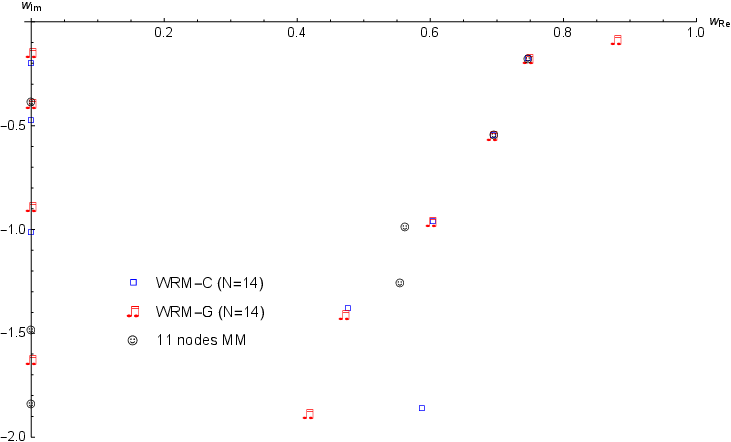}
\caption{The roots of the master equation for axial gravitational QNMs were obtained using the matrix method (MM), the Weighted Residual Method with the collocation method (WRM-C), and the Weighted Residual Method with the Galerkin method (WRM-G), where $L=2$.
}
\lb{Fig7}
\end{figure}

Criterion V: As we all know, each eigenvalue corresponds to an eigenvector (or eigenfunction) of the eigen equation. Therefore, each root from Eq.(\ref{H1}) must correspond to an eigenvector, which can be calculated by substituting the eigenvalue into the eigen equation. It is observed that the physical eigenvectors, representing the wave function of QNMs, exhibit a regular form, while the eigenvectors associated with extraneous roots appear irregular. Moreover, the form of the eigenfunction for extraneous roots is highly sensitive to numerical precision in computations, whereas the physical results remain stable under different levels of numerical precision. In Figure 8, we provide examples to illustrate this property, showing a physical solution $w=0.747343-0.177985i$ (depicted in the left figure) and an extraneous root $w=-0.0790609i$ (depicted in the right figure). We calculate the eigenvectors with 10, 20, and 30 significant digits, respectively. It is evident that the eigenfunction forms for extraneous roots exhibit irregular patterns and vary significantly with different levels of precision. In contrast, the eigenfunction patterns for physical solutions consistently exhibit graceful and stable forms.

\begin{figure*}[htbp]
\centering
\includegraphics[width=0.8\columnwidth]{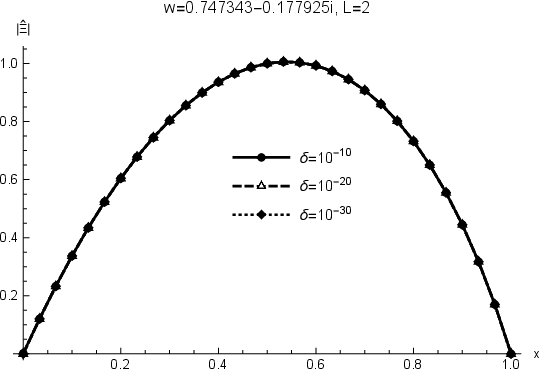}\includegraphics[width=0.8\columnwidth]{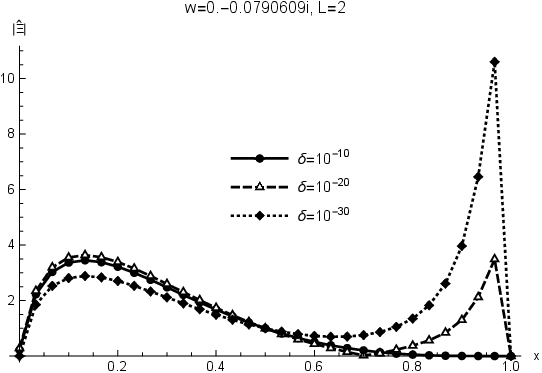}
\caption{The eigenvectors are obtained for the physical solution (left figure) and extraneous roots (right figure) using the matrix method. The calculation is performed with 31 nodes and $L=2$. We define $\Phi(r)=(1-\frac{r_p}{r})^{-1}(\frac{r_p}{r})^{-1}(r-r_p)^{-i\omega r_p}r^{2i\omega r_p}e^{i\omega r}\Xi(r)$, $x=1-\frac{r_p}{r}$, $w=\omega r_p$, and $\hat{\Xi}(x)\equiv\frac{\Xi(x)}{\Xi(\frac{1}{2})}$. In the figure, we present the numerical results of the eigenvectors calculated with $w$ retaining 10, 20, and 30 significant digits (represented respectively by $\delta=10^{-10}$, $10^{-20}$, $10^{-30}$).
}
\lb{Fig8}
\end{figure*}

The five criteria described above help to distinguish the properties of physical solutions and extraneous roots from various perspectives. In general, we can simultaneously apply two or three (or more) criteria to reliably identify the physical root from Eq. (\ref{H1}).

In addition, there is another commonly used but not always effective trick: if we know the QNMs frequency $\omega_A$ of a simpler potential $V_A$, and we want to find the QNMs frequency $\omega_B$ corresponding to a more complicated potential $V_B$, we can construct a  potential function $V=V(\epsilon)$ with a parameter $\epsilon$ such that $V(\epsilon=0)=V_A$ and $V(\epsilon=1)=V_B$. Then, we can divide the parameter $\epsilon$ into many segments, with $\epsilon_k=k\Delta\epsilon$ for $k=0,1,\cdots,N=1/\Delta\alpha$. By using $\epsilon=\epsilon_{k}$ as the initial value to calculate the QNMs frequency at $\epsilon=\epsilon_{k+1}$, we can eventually determine the resonant frequency $\omega_B$ at $\epsilon=1$. If the curve of $\omega=\omega(\epsilon)$ is smooth and continuous, we can roughly consider that the value obtained from this process is a physical solution for $\omega_B$.

\section{Conclusion}
\renewcommand{\theequation}{8.\arabic{equation}} \setcounter{equation}{0}

This paper provides a review of the matrix method and introduces an improved matrix method. We employ the Jordan decomposition to rewrite the square coefficient matrix in the form of Eq.(\ref{MM14}). The invertible matrix $P_K$ and the \textcolor{black}{nilpotent} Jordan matrix $J_K$ in our Jordan decomposition can be expressed using Eq.(\ref{MM11}) and Eq.(\ref{MM13}), respectively. Subsequently, we utilize Eq.(\ref{MM16}) and Eq.(\ref{MM17}) to solve for the frequencies of the QNMs in the gravitational perturbation equations and their corresponding wave functions. Notably, this method eliminates the need for calculating the inverse matrix of the square matrix, thus conserving computational resources and significantly improving efficiency.

On the other hand, we also consider solving the QNMs master equation using the weighted residual method. After the coordinate transformation $x=1-\frac{r_p}{r}$ and utilizing Eq.(\ref{WRM1}), the domain of $x$ is given by $x\in[0,1]$, with the wave function $\Xi$ satisfying the boundary condition Eq.(\ref{WRM2}). By substituting the trial function (\ref{WRM5}) into the master equation and integrating the equation with a weight function $\sigma_k$, the differential equation can be transformed into a matrix equation. Finally, we can calculate the QNMs frequencies using Eq. (\ref{WRM9}) and determine the corresponding wave function. In this paper, we choose the collocation method and Galerkin method as examples to demonstrate the calculation of the weighted residual method. However, it should be noted that there are other weight functions available, such as the least squares method, subdomain method, and method of moments. Furthermore, we can also select a trial function that satisfies the boundary conditions Eq.(\ref{WRM2}) but differs from Eq.(\ref{WRM6}), such as $u_i=\sin(i\pi x)$. In summary, the weighted residual method is a flexible, simple, and highly accurate numerical method for studying QNMs.

Frequency-domain methods, such as CFM, HHM, AIM, MM, WRM, etc., are capable of yielding both physical solutions and extraneous roots simultaneously. Hence, it is important to address the exclusion of extraneous roots. In order to facilitate the selection of physical solutions, we present five criteria.

TABLE I shows that the results computed by MM and WRM exhibit exceptional precision, surpassing that achieved by many well-known methods, such as the WKB approximation. This fact underscores the reliability of these two methods in accurately studying gravitational perturbations within the internal regions of black holes. Further discussion on this case will be presented in the subsequent appendix B.

To clarify, it should be noted that although both MM and WRM in this article use the condition that ``the determinant of the square matrix vanishes" ($\text{det}(M)=0$) as the equation for computing QNMs frequencies, this condition is not unique to MM and WRM. After appropriate rearrangement, many Frequency-Domain methods can also use this condition as an algebraic equation for calculating QNMs frequencies. We will demonstrate this fact in Appendix A.

\textcolor{black}{
This paper demonstrates the feasibility of two methods using the example of a spherically symmetric Schwarzschild black hole. However, for rotating black hole spacetimes, after separating variables in the perturbation equation, two equations arise: the radial equation and the angular equation. They share common eigenvalues $\omega$ and $\lambda$, where $\omega$ represents the QNMs frequency, while $\lambda$ is associated with $L$. Both eigenvalue equations need to be solved simultaneously to obtain the values of $\omega$ and $\lambda$. In \cite{Lin3}, we have demonstrated that this problem in Kerr and Kerr-Sen black hole spacetime can indeed be solved using the (original) matrix method, therefore the two methods proposed in this paper should also be applicable for solving the QNMs problem in rotating black hole spacetimes. We will work on this topic in the future.
}

In fact, the matrix method and weighted residual method can not only solve the QNMs equations for black holes or compact stars, but also allow for the straightforward calculation of perturbation dynamics in other fields. Recently, we have been applying these methods to study atmospheric perturbations in geospace \cite{spacephysics1,spacephysics2}. It is believed that these two numerical methods can significantly contribute to our understanding and description of the physical processes involved.

\appendix

\section{The Matrix Form of Continuous Fraction Method
}
\renewcommand{\theequation}{A.\arabic{equation}} \setcounter{equation}{0}

While MM and WRM can both use the condition that the determinant of the corresponding square matrix vanishes to obtain the QNMs frequencies, it is important to emphasize that ``the determinant vanishes" is not an exclusive condition for MM and the WRM. In fact, many Frequency-Domain methods can be expressed in matrix form and utilize the condition of ``the determinant vanishes" for solving. In this section, taking CFM as an example, we demonstrate the matrix form of this method. To do so, let us consider the axial case of Eq. (\ref{Metric7}):

\bqn
\lb{CFMM1}
&&(x^3-2x^2+x)\Psi''(x)+\left[(3-4iw)x^2 +(8iw-4)x\right.\nb\\
&&\left.+(1-2iw)\right.]\Psi'(x)+\left[-(4w^2+4iw+3)x \right.\nb\\
&&\left.+ (8w^2+4iw-L^2-L+3)\right]\Psi(x)=0.
\eqn
According to CFM, we can set
\bqn
\lb{CFMM2}
\Psi(x)=\sum^\infty_{k=0}a_kx^k,
\eqn
so
\bqn
\lb{CFMM3}
x\Psi(x)&=&\sum^\infty_{k=1}a_{k-1}x^k,\nb\\
\Psi'(x)&=&\sum^\infty_{k=0}(k+1)a_{k+1}x^k,\nb\\
x\Psi'(x)&=&\sum^\infty_{k=0}na_{k}x^k,\nb\\
x^2\Psi(x)&=&\sum^\infty_{k=1}(k-1)a_{k-1}x^k,\nb\\
\Psi''(x)&=&\sum^\infty_{k=0}(k+1)(k+2)a_{k+2}x^k,\nb\\
x\Psi''(x)&=&\sum^\infty_{k=0}k(k+1)a_{k+1}x^k,\nb\\
x^2\Psi''(x)&=&\sum^\infty_{k=0}k(k-1)a_{k}x^k,\nb\\
x^3\Psi''(x)&=&\sum^\infty_{k=1}(k-1)(k-2)a_{k-1}x^k.
\eqn

Substituting the above equation into Eq.(\ref{CFMM1}), we obtain:
\bqn
\lb{CFMM4}
&&\sum^\infty_{k=1}\left[\tilde{\alpha}_k a_{k+1}+\tilde{\beta}_k a_k+\tilde{\gamma}_k a_{k-1}\right]x^k\nb\\
&&+\tilde{\alpha}_0 a_{1}+\tilde{\beta}_0 a_0=0
\eqn
namely
\bqn
\lb{CFMM5}
&&\tilde{\alpha}_k a_{k+1}+\tilde{\beta}_k a_k+\tilde{\gamma}_k a_{k-1}=0~~~~\text{for}~k\ge 1\nb\\
&&\tilde{\alpha}_0 a_{1}+\tilde{\beta}_0 a_0=0
\eqn
where,
\bqn
\lb{CFMM6}
\tilde{\alpha}_k&=&k(k+1)+(1-2iw)(k+1),\nb\\
\tilde{\beta}_k&=&(8w^2+4iw-L^2-L+3)\nb\\
&&+(8iw-4)k-2k(k-1),\nb\\
\tilde{\gamma}_k&=&(k-1)(k-2)+(3-4iw)(k-1)\nb\\
&&-(4w^2+4iw+3).
\eqn
as $k\gg 1$, from the recursive relationship, we have
\bqn
\lb{CFMM7}
\frac{a_{k+1}}{a_k}&\approx&\frac{a_k}{a_{k-1}}\approx\frac{-1\mp\sqrt{1-4\frac{\tilde{\alpha_k}}{\tilde{\beta_k}}\frac{\tilde{\gamma_k}}{\tilde{\beta_k}}}}{2\frac{\tilde{\alpha_k}}{\tilde{\beta_k}}}\nb\\
&=&1\pm\sqrt{\frac{-2iw}{k}}-\frac{1+2iw}{k}+{\cal O}\left(\frac{1}{k^{3/2}}\right)~~~
\eqn
According to the convergence condition, the sign in front of the second term above should be negative. Therefore, we can let $a_0=1$ and then use the above equation as the formula for solving the QNMs frequencies.

On the other hand, if we express Eq.(\ref{CFMM4}) in matrix form, it is given by
\bqn
\lb{CFMM8}
&&\begin{bmatrix}
\tilde{\beta}_0 & \tilde{\alpha}_0 & 0 & 0 & 0 & \cdots  \\
\tilde{\gamma}_1 & \tilde{\beta}_1 & \tilde{\alpha}_1 & 0 & 0 & \cdots \\
0 & \tilde{\gamma}_2 & \tilde{\beta}_2 & \tilde{\alpha}_2 & 0 & \cdots \\
0 & 0 & \tilde{\gamma}_3 & \tilde{\beta}_3 & \tilde{\alpha}_3 & \cdots \\
\cdots & \cdots & \cdots & \cdots & \cdots & \cdots \\
\end{bmatrix}\begin{bmatrix}
a_0 \\
a_1 \\
a_2\\
a_3\\
\cdot\cdot\cdot\\
\end{bmatrix}\nb\\
&&\equiv M_\text{CFM}F_\text{CFM}=0
\eqn
and QNMs frequencies should satisfy:
\bqn
\lb{CFMM9}
\text{det}(M_\text{CFM})=0
\eqn
Therefore, Eq.(\ref{CFMM9}) can also serve as the condition for solving the QNMs frequencies by using CFM.

On the other hand, it is not necessary to require that MM satisfies the condition ``the determinant vanishes" as the formula for solving the QNMs frequencies. For simplicity, we can start from Eq.(\ref{Metric7}) and consider the simplest difference equation:
\bqn
\lb{CFMM10}
&&\text{for}~i=0\nb\\
\Psi'(x_0)&=&\frac{1}{h}\left[-\frac{3}{2}\Psi(x_0)+2\Psi(x_1)-\frac{1}{2}\Psi(x_2)\right]\nb\\
\Psi''(x_0)&=&\frac{1}{h^2}\left[\Psi(x_0)-2\Psi(x_1)+\Psi(x_2)\right]\nb\\
&&\text{for}~n-1\ge i\ge 1\nb\\
\Psi'(x_i)&=&\frac{1}{h}\left[-\frac{1}{2}\Psi(x_{i-1})+\frac{1}{2}\Psi(x_{i+1})\right]\nb\\
\Psi''(x_i)&=&\frac{1}{h^2}\left[\Psi(x_{i-1})-2\Psi(x_{i})+\Psi(x_{i+1})\right]\nb\\
&&\text{for}~i=n\nb\\
\Psi'(x_n)&=&\frac{1}{h}\left[\frac{1}{2}\Psi(x_{n-2})-2\Psi(x_{n-1})+\frac{3}{2}\Psi(x_n)\right]\nb\\
\Psi''(x_n)&=&\frac{1}{h^2}\left[\Psi(x_{n-2})-2\Psi(x_{n-1})+\Psi(x_{n})\right]\nb\\
\eqn
As mentioned before, $h=1/n$ and hence $x_i=i/n$. Therefore, we can express the equations for the QNMs frequencies as a system of algebraic equations:
\bqn
\lb{CFMM11}
&&\hat{\alpha}_i \Psi(x_{i+1})+\hat{\beta}_i \Psi(x_i)+\hat{\gamma}_i \Psi(x_{i-1})=0~\text{for}~0\ge i\ge 1\nb\\
&&\hat{W}_2 \Psi(x_2)+\hat{W}_1 \Psi(x_1)+\hat{W}_0\Psi(x_0)=0
\eqn
as $i=n$, we have
\bqn
\lb{CFMM12}
\hat{A} \Psi(x_{n})+\hat{B} \Psi(x_{n-1})+\hat{C}\Psi(x_{n-2})=0
\eqn
where
\bqn
\lb{CFMM13}
\hat{W}_2&=&\frac{1}{h^2}\alpha_2(x_0)-\frac{1}{2h}\alpha_1(x_0),\nb\\
\hat{W}_1&=&-\frac{2}{h^2}\alpha_2(x_0)+\frac{2}{h}\alpha_1(x_0),\nb\\
\hat{W}_0&=&\frac{1}{h^2}\alpha_2(x_0)-\frac{3}{2h}\alpha_1(x_0)+\alpha_0(x_0)\nb\\
\hat{\alpha}_i&=&\frac{1}{h^2}\alpha_2(x_i)+\frac{1}{2h}\alpha_1(x_i),\nb\\
\hat{\beta}_i&=&-\frac{2}{h^2}\alpha_2(x_i)+\alpha_0(x_i),\nb\\
\hat{\gamma}_i&=&\frac{1}{h^2}\alpha_2(x_i)-\frac{1}{2h}\alpha_1(x_i)\nb\\
\hat{A}&=&\frac{1}{h^2}\alpha_2(x_n)+\frac{3}{2h}\alpha_1(x_n)+\alpha_0(x_n),\nb\\
\hat{B}&=&-\frac{2}{h^2}\alpha_2(x_n)-\frac{2}{h}\alpha_1(x_n),\nb\\
\hat{C}&=&\frac{1}{h^2}\alpha_2(x_n)+\frac{1}{2h}\alpha_1(x_n),
\eqn
and $\alpha(x_j)\equiv\alpha(x=x_j,\omega)$.

Expressing it in matrix form, we have:
\bqn
\lb{CFMM14}
&&\begin{bmatrix}
\hat{W}_0 & \hat{W}_1 & \hat{W}_2 & 0 & 0 & \cdots  \\
\hat{\gamma}_1 & \hat{\beta}_1 & \hat{\alpha}_1 & 0 & 0 & \cdots \\
0 & \hat{\gamma}_2 & \hat{\beta}_2 & \hat{\alpha}_2 & 0 & \cdots \\
0 & 0 & \hat{\gamma}_3 & \hat{\beta}_3 & \hat{\alpha}_3 & \cdots \\
\cdots & \cdots & \cdots & \cdots & \cdots & \cdots \\
\cdots & 0 & \hat{\gamma}_{n-2} & \hat{\beta}_{n-2} & \hat{\alpha}_{n-2} & 0 \\
\cdots & 0 & 0 & \hat{\gamma}_{n-1} & \hat{\beta}_{n-1} & \hat{\alpha}_{n-1} \\
\cdots & 0 & 0 & \hat{C} & \hat{B} & \hat{A} \\
\end{bmatrix}\begin{bmatrix}
\Psi(x_0) \\
\Psi(x_1) \\
\Psi(x_2)\\
\Psi(x_3)\\
\cdot\cdot\cdot\\
\Psi(x_{n-2})\\
\Psi(x_{n-1})\\
\Psi(x_n)\\
\end{bmatrix}\nb\\
&&\equiv M_\text{MM}F_\text{MM}=0
\eqn
In matrix method, We can calculate the QNMs frequencies by following relation
\bqn
\lb{CFMM15}
\text{det}(M_\text{MM})=0.
\eqn

We can also employ a similar method to the Continued Fraction Method (CFM) to solve the QNMs frequencies. From Eq(\ref{CFMM11}), we let $\Psi(x_0) = 1$, so
\bqn
\lb{CFMM16}
\Psi(x_1)&=&\frac{\hat{W}_2\hat{\gamma}_1-\hat{W}_0\hat{\alpha}_1}{\hat{W}_1\hat{\alpha}_1-\hat{W}_2\hat{\beta}_1}\nb\\
\Psi(x_2)&=&\frac{\hat{W}_1\hat{\gamma}_1-\hat{W}_0\hat{\beta}_1}{\hat{W}_1\hat{\alpha}_1-\hat{W}_2\hat{\beta}_1}\nb\\
\Psi(x_{i+1})&=&-\frac{\hat{\beta}_i}{\hat{\alpha}_i} \Psi(x_i)-\frac{\hat{\gamma}_i}{\hat{\alpha}_i} \Psi(x_{i-1})~~~(i\ge 2)\nb\\
\eqn
Therefore, we can calculate $\Psi(x_{n-2})$, $\Psi(x_{n-1})$, and $\Psi(x_{n})$ with undetermined parameter $w$, namely the QNMs frequency. Next, by substituting above three solutions into Eq.(\ref{CFMM12}), we can get $w$.

Let's consider the case of axial perturbation as an example, where we calculate the QNMs frequency for $L=0$. By using the CFM, we set $\Psi(x)=\sum_{k=0}^{15}a_k x^k$, and by applying Eq. (\ref{CFMM7}), it is given by
\bqn
\lb{CFMM17}
\left.\frac{a_{k+1}}{a_k}\right|_{k=14}=\left.1-\sqrt{\frac{-2iw}{k}}-\frac{1+2iw}{k}\right|_{k=14}
\eqn
so the QNMs frequency with overtone $\bar{n}=0$ $w=w_\text{CFM-A}=0.7473434579 - 0.1779236053i$.

Let's set the order of the matrix $M_\text{CFM}$ to be 15 and compute using the formula (\ref{CFMM9}), we can obtain $w=w_\text{CFM-B}=0.7473398449 - 0.1779208062i$. The difference between $w_\text{CFM-B}$ and $w_\text{CFM-A}$ is due to the truncation errors from different mathematical methods.

If we solve the problem by MM with 15-nodes. From Eq.(\ref{CFMM12}) or Eq.(\ref{CFMM15}) respectively, we can get the same results: $w=w_\text{MM}=0.7459970733 - 0.1784093416i$.

From the discussion in this section, we have demonstrated that the condition for ``the determinant vanishes" is not exclusive to the matrix method and the weighted residual method. Frequency-Domain methods, for example the continued fraction method, can also utilize this condition to calculate QNMs frequencies. On the other hand, Sometimes, the matrix method and the weighted residual method can solve eigenvalues without relying on the condition of ``the determinant vanishes".

From the above results, we also observe that when using the continued fraction method for calculations, we need to first obtain the recurrence relations (\ref{CFMM4}) and (\ref{CFMM5}). This requires the function $\alpha_i(x,w)$ in Eq.(\ref{Metric7}) to be the simple rational expression. However, if $\alpha_i(x,w)$ has a complex form or even numerical expression, the continued fraction method cannot obtain the recurrence relations, greatly affecting the efficiency of this method. On the other hand, the advantages of the matrix method and weighted residual method lie in their ability to directly utilize mathematical software such as Mathematica or MATLAB for calculations, without the need to derive recurrence relations.

\section{Gravitational perturbations in the internal region of a Schwarzschild black hole
}
\renewcommand{\theequation}{B.\arabic{equation}} \setcounter{equation}{0}

In general relativity, it is impossible to propagate any information from the interior region of a black hole to the exterior region. This fact means that there are no chances to observe and verify any physical effects predicted by theoretical research within the interior of a black hole. Nevertheless, the interior region of a black hole exhibits many exotic properties, including singularities and an interchange between radial space and time components. These exotic properties pose a formidable challenge to existing physical concepts and have piqued the curiosity of researchers. The proposal of high-precision frequency-domain methods has provided us with the means to compute and demonstrate the dynamical processes of perturbations within the event horizon of black holes. Therefore, in the appendix, we aim to study the dynamics of gravitational perturbations in the interior region of black holes using the matrix method.

According to Equation (\ref{Metric1}) and Equation (\ref{Metric2}), within the spacetime inside the event horizon of a black hole, all particles move towards the singularity at its center. However, the potential becomes divergent at the center, requiring the \textcolor{black}{gravitational perturbation} equation function to vanish there. Therefore, the boundary conditions within the black hole can be expressed as follows:
\bqn
\lb{A1}
\Phi(r\rightarrow r_p^-)&\propto& (r_p-r)^{-i\omega r_p},\nb\\
\Phi(r\rightarrow 0)&\rightarrow& 0.
\eqn
To cancel the irregularity at the horizon, we can define the QNM equation function as $\Phi=(r_p-r)^{-i\omega r_p}\Theta(r)$. Here, $\Theta(r\rightarrow r_p^-)$ approaches a constant value, and $\Theta(r\rightarrow 0)$ tends to zero.

The matrix method requires the coordinate transformation $y=\frac{r}{r_p}$ and $w=\omega r_p$, so that $y\in[0,1]$ and the QNM equation inside the black hole becomes:
\bqn
\lb{A2}
&&\text{For axial case:}\nb\\
&&(3 - L y - L^2 y + i w y + w^2 y^2 (1 + y)) \Theta(y) \nb\\
&&+ (y - 2 i w y^2) \Theta'(y) - (1 - y) y^2 \Theta''(y)=0,\nb\\
&&\text{For polar case:}\nb\\
&&\{i w y [3 - (2 - L - L^2) y]^2-9 + 9 (2 - L - L^2) y\nb\\
&& - 3 (2 - L - L^2)^2 y^2 +
    L (1 + L) (2 - L - L^2)^2 y^3\nb\\
&& + w^2 y^2 (1 + y) (3 + (-2 + L + L^2) y)^2\} \Theta(y)\nb\\
&& + \{y [3 - (2 - L - L^2) y]^2 \nb\\
&&- 2 i w y^2 (3 - (2 - L - L^2) y)^2\} \Theta'(y) \nb\\
&&- (1 - y) y^2 (3 - (2 - L - L^2) y)^2 \Theta''(y)=0.
\eqn
In Eq. [\ref{A2}], the wave function $\Theta$ is regular at the boundary points $y=0,1$. We deliberately write the coefficients of the wave function and its first and second order derivatives in these two equations without denominators to avoid the problem of the coefficients' denominator vanishing. Next, we present the calculation results in Figure 9 and Figure 10.

\begin{figure}[htbp]
\centering
\includegraphics[width=0.8\columnwidth]{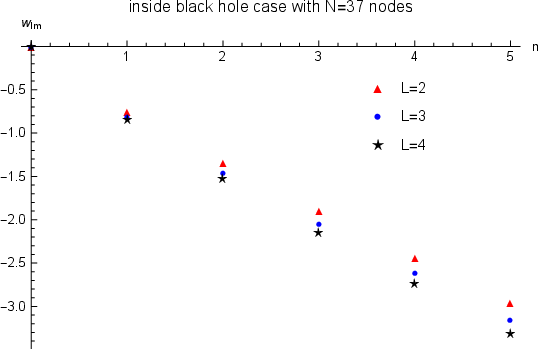}
\caption{The gravitational perturbation frequency in the internal region of the Schwarzschild black hole, obtained by the matrix method with 37 nodes
}
\lb{Fig9}
\end{figure}

We calculated the \textcolor{black}{gravitational perturbation} frequencies in the interior region of the Schwarzschild black hole, and it was found that the real part of the \textcolor{black}{gravitational perturbation} frequency vanishes. Therefore, we only plot the imaginary part of the \textcolor{black}{gravitational perturbation} frequency $w_\text{Im}$ with a lower overtone number $n$ in Figure 9.

\begin{figure*}[htbp]
\centering
\includegraphics[width=0.7\columnwidth]{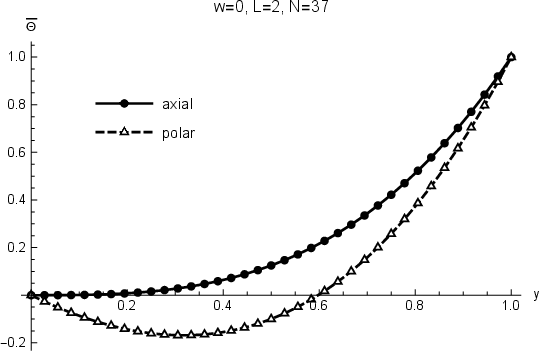}\includegraphics[width=0.7\columnwidth]{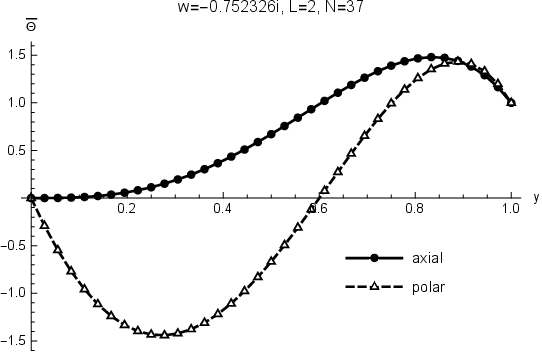}\includegraphics[width=0.7\columnwidth]{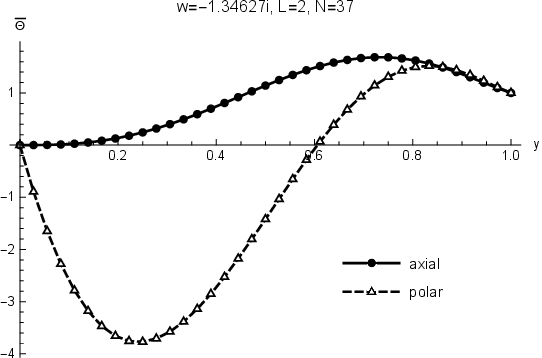}
\includegraphics[width=0.7\columnwidth]{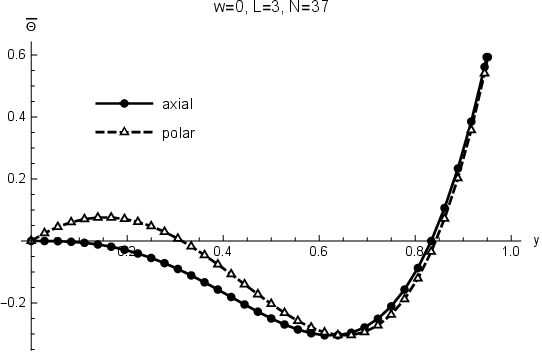}\includegraphics[width=0.7\columnwidth]{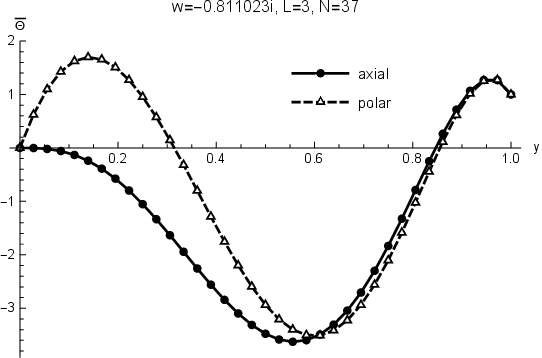}\includegraphics[width=0.7\columnwidth]{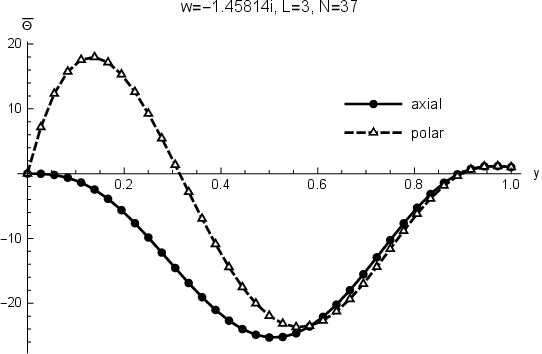}
\includegraphics[width=0.7\columnwidth]{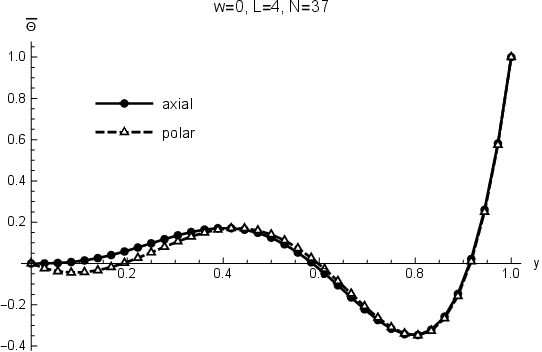}\includegraphics[width=0.7\columnwidth]{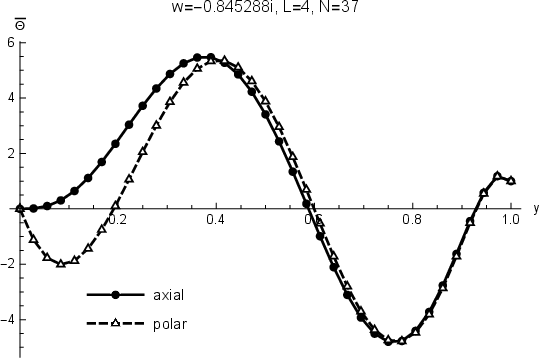}\includegraphics[width=0.7\columnwidth]{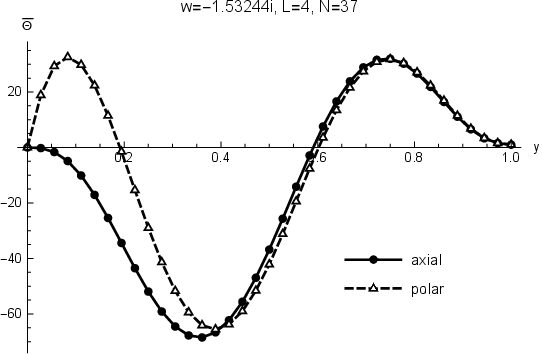}
\caption{The \textcolor{black}{gravitational perturbation} frequency and the wave functions in the internal region of the Schwarzschild black hole are obtained by the matrix method with 37 nodes, where $\Phi(r)=(r_p-r)^{-i\omega r_p}\Theta(r)$, $\bar{\Theta}(r)\equiv\frac{\Theta(r)}{\Theta(\infty)}$, $y=\frac{r}{r_p}$ and $w=\omega r_p$.
}
\lb{Fig10}
\end{figure*}

Figure 10 shows the corresponding axial and polar wave functions at \textcolor{black}{gravitational perturbation} frequencies for $n=0,1,2$ and $L=2,3,4$ in the interior case. By using the relation $\Phi(r)=(r_p-r)^{-i\omega r_p}\Theta(r)$, we establish the relation between $\Theta$ and $y=\frac{r}{r_p}$. Since the \textcolor{black}{gravitational perturbation} frequencies in this case only have imaginary parts, the corresponding wave functions also have only real parts. For wave functions similar to \textcolor{black}{gravitational perturbation} QNMs outside black holes, although the \textcolor{black}{gravitational perturbation} frequencies are the same, the wave functions differ between the axial and polar cases.

The absence of a real part in the \textcolor{black}{gravitational perturbation} frequencies in the interior region of a black hole implies that the gravitational perturbation exponentially decreases without oscillation inside the black hole, indicating that waves and particles inside the black hole are inevitably attracted to the central singularity. This property supports the conclusions of the physical theory of black holes. Furthermore, it is worth noting that the $n=0$ \textcolor{black}{gravitational perturbation} frequencies almost vanish, but its corresponding wave function is not zero. This implies that the perturbation of this mode can persist in this state for a long time. Therefore, in addition to the central singularity point, the spacetime inside the Schwarzschild black hole may exhibit some tiny quasi-structural features.

In fact, although the information inside a black hole cannot cross the event horizon and reach the outside, studying the gravitational perturbations inside the black hole is still meaningful. We believe that perturbations exist everywhere and at any time in the real physical world, including inside a black hole. If the spacetime inside a black hole is unstable under perturbation, it may suggest that this spacetime is not physically real. This criterion can be applied to examine the internal structure of various modified gravity and quantum gravity black holes, as well as different types of regular black holes. If these black holes are unstable under perturbation, their solutions may not have physical relevance in the real universe. The authors plan to conduct further investigations in the future.

\begin{acknowledgments}
This project was supported by
National Natural Science Foundation of China (NNSFC) under contract No. 42230207 and the Fundamental Research Funds for the Central Universities, China University of Geosciences (Wuhan) with No. G1323523064.
\end{acknowledgments}

\end{document}